\documentclass[twocolumn, tighten]{aastex62}
\usepackage{amsmath}
\usepackage[caption=false]{subfig}

\shorttitle{DSHARP III: Spirals in the Elias 27, IM Lup, and WaOph 6 Disks}
\shortauthors{Huang et al.}

\begin{document}

\title{The Disk Substructures at High Angular Resolution Project (DSHARP): \\ 
III. Spiral Structures in the Millimeter Continuum of the Elias 27, IM Lup, and WaOph 6 Disks}

\correspondingauthor{Jane Huang}
\email{jane.huang@cfa.harvard.edu}

\author{Jane Huang}
\affiliation{Harvard-Smithsonian Center for Astrophysics, 60 Garden Street, Cambridge, MA 02138, United States of America}

\author{Sean M. Andrews}
\affiliation{Harvard-Smithsonian Center for Astrophysics, 60 Garden Street, Cambridge, MA 02138, United States of America}

\author{Laura M. P\'erez}
\affiliation{Departamento de Astronom\'ia, Universidad de Chile, Camino El Observatorio 1515, Las Condes, Santiago, Chile}

\author{Zhaohuan Zhu}
\affiliation{Department of Physics and Astronomy, University of Nevada, Las Vegas, 4505 S. Maryland Pkwy, Las Vegas, NV, 89154, USA}

\author{Cornelis P. Dullemond}
\affiliation{Zentrum f\"ur Astronomie, Heidelberg University, Albert Ueberle Str. 2, 69120 Heidelberg, Germany}

\author{Andrea Isella}
\affiliation{Department of Physics and Astronomy, Rice University, 6100 Main Street, Houston, TX 77005, United States of America}

\author{Myriam Benisty}
\affiliation{Unidad Mixta Internacional Franco-Chilena de Astronom\'{i}a, CNRS/INSU UMI 3386, Departamento de Astronom\'ia, Universidad de Chile, Camino El Observatorio 1515, Las Condes, Santiago, Chile}
\affiliation{Univ. Grenoble Alpes, CNRS, IPAG, 38000 Grenoble, France}

\author{Xue-Ning Bai}
\affiliation{Institute for Advanced Study and Tsinghua Center for Astrophysics, Tsinghua University, Beijing 100084, China}

\author{Tilman Birnstiel}
\affiliation{University Observatory, Faculty of Physics, Ludwig-Maximilians-Universit\"at M\"unchen, Scheinerstr. 1, 81679 Munich, Germany}

\author{John M. Carpenter}
\affiliation{Joint ALMA Observatory, Avenida Alonso de C\'ordova 3107, Vitacura, Santiago, Chile}

\author{Viviana V. Guzm\'an}
\affiliation{Joint ALMA Observatory, Avenida Alonso de C\'ordova 3107, Vitacura, Santiago, Chile}
\affiliation{Instituto de Astrof\'isica, Pontificia Universidad Cat\'olica de Chile, Av. Vicu\~na Mackenna 4860, 7820436 Macul, Santiago, Chile}

\author{A. Meredith Hughes}
\affiliation{Department of Astronomy, Van Vleck Observatory, Wesleyan University, 96 Foss Hill Drive, Middletown, CT 06459, USA}

\author{Karin I. \"Oberg}
\affiliation{Harvard-Smithsonian Center for Astrophysics, 60 Garden Street, Cambridge, MA 02138, United States of America}

\author{Luca Ricci}
\affiliation{Department of Physics and Astronomy, California State University Northridge, 18111 Nordhoff Street, Northridge, CA 91130, USA}

\author{David J. Wilner}
\affiliation{Harvard-Smithsonian Center for Astrophysics, 60 Garden Street, Cambridge, MA 02138, United States of America}

\author{Shangjia Zhang}
\affiliation{Department of Physics and Astronomy, University of Nevada, Las Vegas, 4505 S. Maryland Pkwy, Las Vegas, NV, 89154, USA}

\begin{abstract}
We present an analysis of ALMA 1.25 millimeter continuum observations of spiral structures in three protoplanetary disks from the Disk Substructures at High Angular Resolution Project. The disks around Elias 27, IM Lup, and WaOph 6 were observed at a resolution of $\sim40-60$ mas ($\sim6-7$ au). All three disks feature $m=2$ spiral patterns in conjunction with annular substructures. Gas kinematics established by $^{12}$CO $J=2-1$ observations indicate that the continuum spiral arms are trailing. The arm-interarm intensity contrasts are modest, typically less than 3. The Elias 27 spiral pattern extends throughout much of the disk, and the arms intersect the gap at $R\sim69$ au. The spiral pattern in the IM Lup disk is particularly complex\textemdash it extends about halfway radially through the disk, exhibiting pitch angle variations with radius and interarm features that may be part of ring substructures or spiral arm branches. Spiral arms also extend most of the way through the WaOph 6 disk, but the source overall is much more compact than the other two disks. We discuss possible origins for the spiral structures, including gravitational instability and density waves induced by a stellar or planetary companion. Unlike the millimeter continuum counterparts of many of the disks with spiral arms detected in scattered light, these three sources do not feature high-contrast crescent-like asymmetries or large ($R>20$ au) emission cavities. This difference may point to multiple spiral formation mechanisms operating in disks. 
\end{abstract}

\keywords{protoplanetary disks---ISM: dust---techniques: high angular resolution}

\section{Introduction} \label{sec:intro}
Spiral arms are thought to provide key clues to the dynamical evolution of protoplanetary disks, although the precise nature of spiral formation mechanisms is highly debated. Proposed mechanisms for inducing spiral features include perturbations by a stellar or planetary companion \citep[e.g.,][]{1979ApJ...233..857G,1986ApJ...307..395L,2002ApJ...565.1257T}, gravitational instability \citep[e.g.,][]{1998ApJ...503..923B, 2004ApJ...609.1045M, 2004MNRAS.351..630L}, and pressure variations due to shadowing from a misaligned inner disk \citep[e.g.,][]{2016ApJ...823L...8M, 2018MNRAS.475L..35M}.

 \begin{deluxetable*}{cccccccc}
\tablecaption{Imaging summary\label{tab:observations}}
 \tabletypesize{\scriptsize}
\colnumbers
\tablehead{
\colhead{Image}&\colhead{Scales}&\colhead{Briggs}&\colhead{Taper}&\colhead{Synthesized beam}&\colhead{$\theta_\text{MRS}$}&\colhead{Peak $I_\nu$}&\colhead{RMS noise}\\
&($''$)&&(mas $\times$ mas ($^\circ$))&(mas $\times$ mas ($^\circ$))&($''$)&(mJy beam$^{-1}$)&(mJy beam$^{-1}$)}
\startdata
Elias 27 continuum (fiducial)&0, 0.06, 0.15, 0.3, 0.6&0.5& $40 \times 20$ $(173)$ & $49 \times 47$ $(47)$ &11, 1.4&4.8&0.014\\
Elias 27 continuum (mid-res)&0, 0.075, 0.15, 0.3, 0.4, 0.6&1.0&$60 \times 40$ $(-6)$&$85\times81$ $(-75)$&11, 1.4&9.4&0.019\\
IM Lup continuum & 0, 0.03, 0.15, 0.45, 0.9, 1.35 & 0.5 & $33 \times 26 $ (138) & $44\times 43$ (115)&11, 3.4 &7.1&0.014\\
WaOph 6 continuum & 0, 0.015, 0.06, 0.18, 0.36, 0.72& 0 & $55\times 10$ (10)&$58\times54$ (84) &10, 1.4& 8.7 &0.017 \\
Elias 27 $^{12}$CO & 0, 0.1, 0.3, 1, 2, 3&1.0 &$100\times 70$ $(-35)$&$132\times111$ (123)& 6.2\tablenotemark{a}, 1.2&44.9 &1.6\\
IM Lup $^{12}$CO & 0, 0.1, 0.3, 1, 2, 3 & 0 & $100\times 100$ & $122\times 115$ (47)& 10, 2.5 &34.4&1.9\\
WaOph 6 $^{12}$CO&0, 0.1, 0.25, 0.75, 1.5 &$0.5$&$100\times 30$ (17)& $126\times 115$ (100) & 11, 1.5 &41.0 &1.3\\
\enddata
\tablecomments{Column descriptions: (1) Image used in this work. (2) Scales used for multi-scale \texttt{CLEAN}. (3) Briggs weighting parameter. (4) Gaussian taper applied during imaging. (5) Synthesized beam. (6) Maximum recoverable scale (MRS). The first value lists the MRS defined by the shortest baseline $L_\text{min}$: $\theta_\text{MRS}\approx\frac{0.6\lambda}{L_\text{min}}$ (radians), where $\lambda$ is the wavelength of the observations. The second value lists the more conservative value of the MRS defined by the 5th percentile of the $uv$ distances, $L_5$: $\theta_\text{MRS}\approx\frac{0.983\lambda}{L_5}$ (radians). See \url{https://almascience.nrao.edu/documents-and-tools/cycle5/alma-technical-handbook/view} for details. The MRS is slightly different for $^{12}$CO and continuum observations because the latter include continuum-only spectral windows. (7) Peak intensity. For $^{12}$CO, this is measured from an image cube with $\Delta v=0.35$ km s$^{-1}$. (8) RMS noise in the image. For $^{12}$CO, this is the per-channel rms measured from an image cube with $\Delta v=0.35$ km s$^{-1}$.}
\tablenotetext{a}{Imaging of Elias 27 $^{12}$CO only includes baselines exceeding 20 k$\lambda$ in order to filter out foreground cloud emission.}
\end{deluxetable*}

Disk spiral structures have primarily been mapped in scattered light observations, which trace small grains in the upper layers of disks \citep[e.g.,][]{2004ApJ...605L..53F, 2012ApJ...748L..22M, 2013ApJ...762...48G, 2018ApJ...862..103D}. However, scattered light does not probe the midplane, where planets are thought to form. Spiral arms have been detected in the $J=2-1$ and $J=3-2$ transitions of $^{12}$CO and the $J=3-2$ transition of $^{13}$CO in a small number of disks, but these lines are  optically thick and thus only probe the disk upper layers \citep[e.g.,][]{2012AA...547A..84T, 2014ApJ...785L..12C, 2017ApJ...840...32T, 2018ApJ...853..162B}. Because of its relatively low optical depth, millimeter continuum emission is crucial for tracing surface density variations in the midplane, and therefore for examining how planetary companions might affect their immediate disk environment. 

Until recently, the disks around Elias 27 and MWC 758 were the only two confirmed cases of spiral features detected in millimeter continuum emission \citep{2016Sci...353.1519P, 2018ApJ...853..162B, 2018ApJ...860..124D}. Hence, little was known about the circumstances under which spiral arms manifest in millimeter continuum emission and about the range of possible morphologies. To characterize disk substructures in a homogeneous fashion, the Disk Substructures at High Angular Resolution Project (DSHARP) undertook a high angular resolution ALMA survey of 20 Class II sources \citep{2018ApJ...869L..41A}. While annular substructures are the most common type of millimeter continuum disk feature detected in this survey, spiral arms are observed in five of these disks (Elias 27, IM Lup, WaOph 6, HT Lup A, and AS 205 N), which brings the overall number of disks known to have millimeter continuum spiral arms to six. HT Lup A and AS 205 N are discussed separately in \citet{2018ApJ...869L..44K} because they belong to multiple disk systems where dynamical interactions between the individual components are likely to induce spiral structures. The present paper focuses on spiral structures detected in three systems without known companions: Elias 27, IM Lup, and WaOph 6. Section \ref{sec:observations} provides an overview of the targets and the observations. Section \ref{sec:features} analyzes the spiral properties. Section \ref{sec:discussion} considers the results in the context of other spiral arm observations and discusses possible origins for these features. Section \ref{sec:summary} summarizes the findings. 

\section{Source properties and observations} \label{sec:observations}

Elias 27 is a 0.8 Myr M0 star located $116\substack{+19\\-10}$ pc away in the $\rho$ Oph star-forming region \citep{luhman99b, 2018AA...616A...1G}. Its disk was the first in which spiral arms were detected in millimeter continuum emission \citep{2016Sci...353.1519P}. IM Lup is a 0.5 Myr K5 star located $158\pm3$ pc away in the Lupus II cloud \citep{alcala17, 2018AA...616A...1G}. Both the gas disk as traced by $^{12}$CO and dust disk as traced by millimeter continuum emission and scattered light are unusually large, stretching out to radii of hundreds of au \citep[e.g.,][]{2008AA...489..633P, 2016ApJ...832..110C, 2018ApJ...863...44A}. Finally, WaOph 6 is a 0.3 Myr K6 star located $123\pm2$ pc away in $\rho$ Oph. \citep{eisner05, 2018AA...616A...1G}. 

The 1.25 mm continuum and $^{12}$CO $J=2-1$ emission of these three disks were observed  as part of DSHARP. A detailed overview of the survey, including the observational setup, calibration, imaging, and rationale, is provided in \citet{2018ApJ...869L..41A}. Unless otherwise specified, the analysis in this work is based on the ``fiducial'' images from the overview paper. In brief, the data were first calibrated with the ALMA pipeline. We subsequently applied phase and amplitude self-calibration to each source, then used multi-scale \texttt{CLEAN} to image the continuum and $^{12}$CO. Tapering and Briggs weighting parameters were selected to balance reduction of beam elongation and PSF sidelobe levels against degradation of resolution or sensitivity. We applied a larger taper for $^{12}$CO compared to the continuum in order to improve sensitivity to the larger emission scales of the line data, which led to larger synthesized beams. In addition, the $^{12}$CO line in the Elias 27 disk was imaged only with baselines longer than 20 k$\lambda$ in order to filter out foreground cloud emission. Imaging parameters and characteristics are listed in Table \ref{tab:observations}.

\begin{figure*}[htp]
\begin{center}
\subfloat[\textit{Top}: ALMA 1.25 mm continuum images of the Elias 27, IM Lup, and WaOph 6 disks. The synthesized beam is shown in the lower left corner of each panel. \textit{Bottom:} The continuum emission deprojected and replotted as a function of disk radius and polar angle.\label{fig:overview}]{\includegraphics[scale = 1]{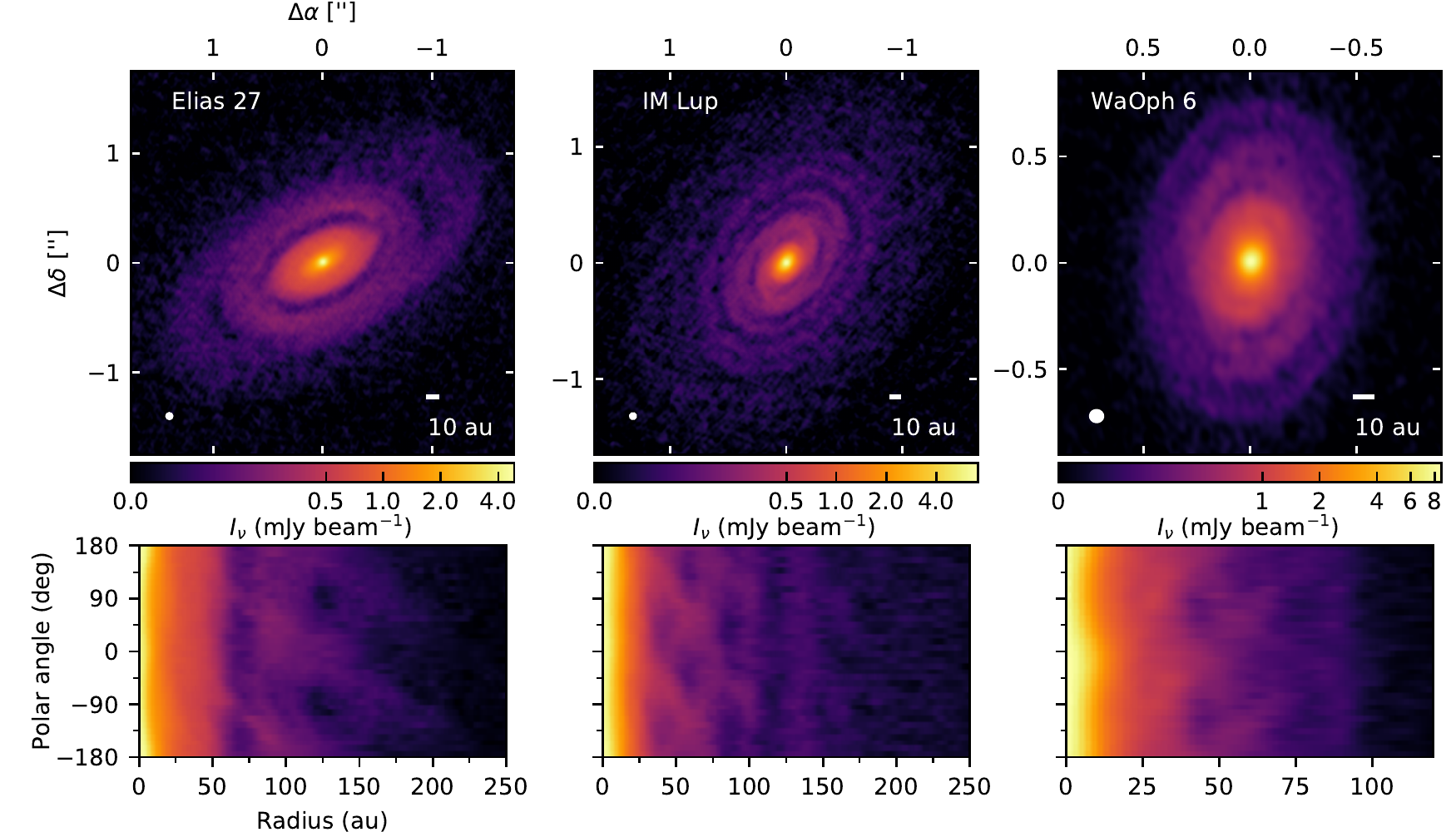}}
\end{center}
\begin{center}
\subfloat[\textit{Top}: Residual emission after subtracting the median radial intensity profile. White stars mark the continuum emission peaks. Ellipses mark the locations of all annular substructures identified in \citet{2018ApJ...869L..42H}. Dotted ellipses correspond to D69 in the Elias 27 disk, D117 in the IM Lup disk, and D79 in the WaOph 6 disk. Solid ellipses correspond to B86 in the Elias 27 disk, B134 in the IM Lup disk, and B88 in the WaOph 6 disk. \textit{Bottom:} Residual emission replotted as a function of disk radius and polar angle. Dotted lines mark the locations of gaps and solid lines mark the locations of bright rings.\label{fig:subtractedall} ]{\includegraphics[scale = 1]{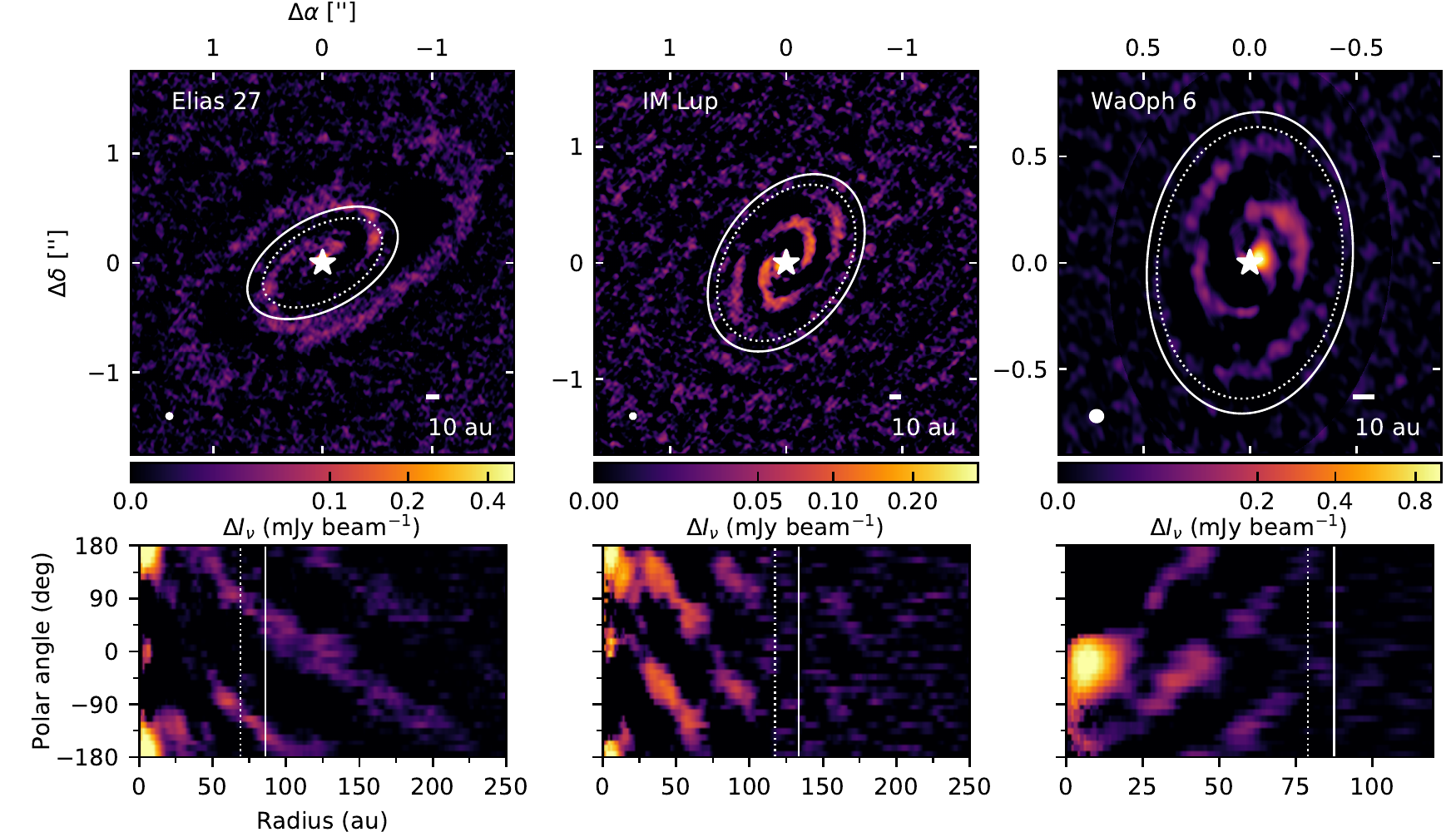}}
\end{center}
\caption{\label{fig:fig1}}
\end{figure*}

\section{Disk features\label{sec:features}}

\subsection{Overview of spiral morphology\label{sec:overview}}

Figure \ref{fig:overview} shows the fiducial ALMA 1.25 mm continuum images of the Elias 27, IM Lup, and WaOph 6 disks from \citet{2018ApJ...869L..41A} as well as maps of the continuum intensity replotted as a function of polar angle and radius in deprojected coordinates. The inclinations and position angles used for deprojection are taken from \citet{2018ApJ...869L..42H} (the Elias 27 disk has a P.A. of 118$\fdg$8 and inclination of 56$\fdg$2, the IM Lup disk has a P.A. of 144$\fdg$5 and inclination of 47$\fdg5$, and the WaOph 6 disk has a P.A. of 174$\fdg$2 and inclination of 47$\fdg$3). The positive $y$-axis of the deprojected coordinate system is rotated east of north by the position angle of the disk, and the polar angle increases in the clockwise direction (i.e., if the position angle of a disk is zero, then $\theta = 90^\circ$ is in the direction of increasing declination and $\theta = 0^\circ$ is in the direction of increasing right ascension). See the Appendix of \citet{2018ApJ...869L..42H} for a diagram. 

In the polar plots, annular substructures (i.e., gaps and rings) appear as vertical bands. These substructures are discussed in detail in \citet{2018ApJ...869L..42H}. We use the same nomenclature to refer to these structures, e.g., D69 refers to a gap at a radius of 69 au, and B86 refers to a bright ring at a radius of 86 au. The spiral structures manifest as emission bands crossing the polar plots diagonally. The spiral patterns can be viewed more readily by subtracting an axisymmetric intensity profile from the disk image (Figure \ref{fig:subtractedall}). For brevity, the resulting plots will be referred to as ``non-axisymmetric residual plots.'' The axisymmetric profile is derived by deprojecting the disk, binning the pixels in au-wide annuli, and finding the median intensity in each bin. 

The three disks are dominated by an $m=2$ spiral pattern (i.e., two-fold rotational symmetry). The spiral pattern in the Elias 27 disk extends from $R\sim50-230$ au, nearly the full extent of the detected millimeter continuum emission. Based on the polar plot in Figure \ref{fig:subtractedall}, the pattern wraps at least $250^\circ$ around the disk. The new high-resolution observations demonstrate that the spiral pattern extends several tens of au further inward than what was visible in previous observations by \citet{2016Sci...353.1519P}. In particular, it can now be seen that the spiral arms intersect with D69, as shown in the leftmost columns of Figure \ref{fig:fig1}. 

Although the overall radial extent of the IM Lup millimeter continuum is comparable to that of Elias 27, the IM Lup spiral pattern is confined to a more compact region extending from $R\sim 25 - 110$ au. The spiral pattern wraps $\sim270^\circ$ around the disk. While the spiral arms largely seem confined within the gap/ring pair D118/B134, there appears to be additional substructure just exterior to B134 that could originate from another pair of spiral arms, a continuation of the interior spiral arms, or a ring that is not well-resolved along the minor axis. Observations with better angular resolution and SNR will be necessary to confirm their nature. 

Both the overall continuum emission and the spiral pattern are more compact for the WaOph 6 disk compared to the other two disks. The spiral pattern is visible from $R\sim25-75$ au and has an azimuthal extent of $\sim215^\circ$. Similarly to IM Lup, the WaOph 6 spirals are nested inside the gap/ring pair D79/B88. 

The main spiral arms may extend further inward and outward than noted for these three disks. Spirals in the outermost regions of the disk are difficult to measure due to low SNR. In the inner few tens of au, spiral arms may also be difficult to detect due to distortions imposed by the PSF, inadequate angular resolution, or insufficient intensity contrasts due to high optical depth. 

The emission bands tracing the spiral arms in the polar plots from Figure \ref{fig:subtractedall} have discontinuities at $R\sim75$ au for IM Lup and $R\sim50$ au for WaOph 6. The non-axisymmetric residual plot for Elias 27 also exhibits a slight radial depression at $R\sim110$ au. These discontinuities correspond to the locations in the polar plots from Figure \ref{fig:overview} where there appears to be additional bright emission between the main spiral arms. This could occur if a pair of spiral arms crosses a bright emission ring or two pairs of spiral arms are separated by a bright emission ring.  Another possibility is that the inter-arm emission arises from ``spurs'' off the main spiral arms, similar to those observed sometimes in spiral galaxies \citep[e.g.,][]{1980ApJ...242..528E}. See Section \ref{sec:pitchangles} for further discussion. 

\subsection{Orientation of the spiral arms}
\begin{figure*}[htp]
\centering
\includegraphics[width=\linewidth]{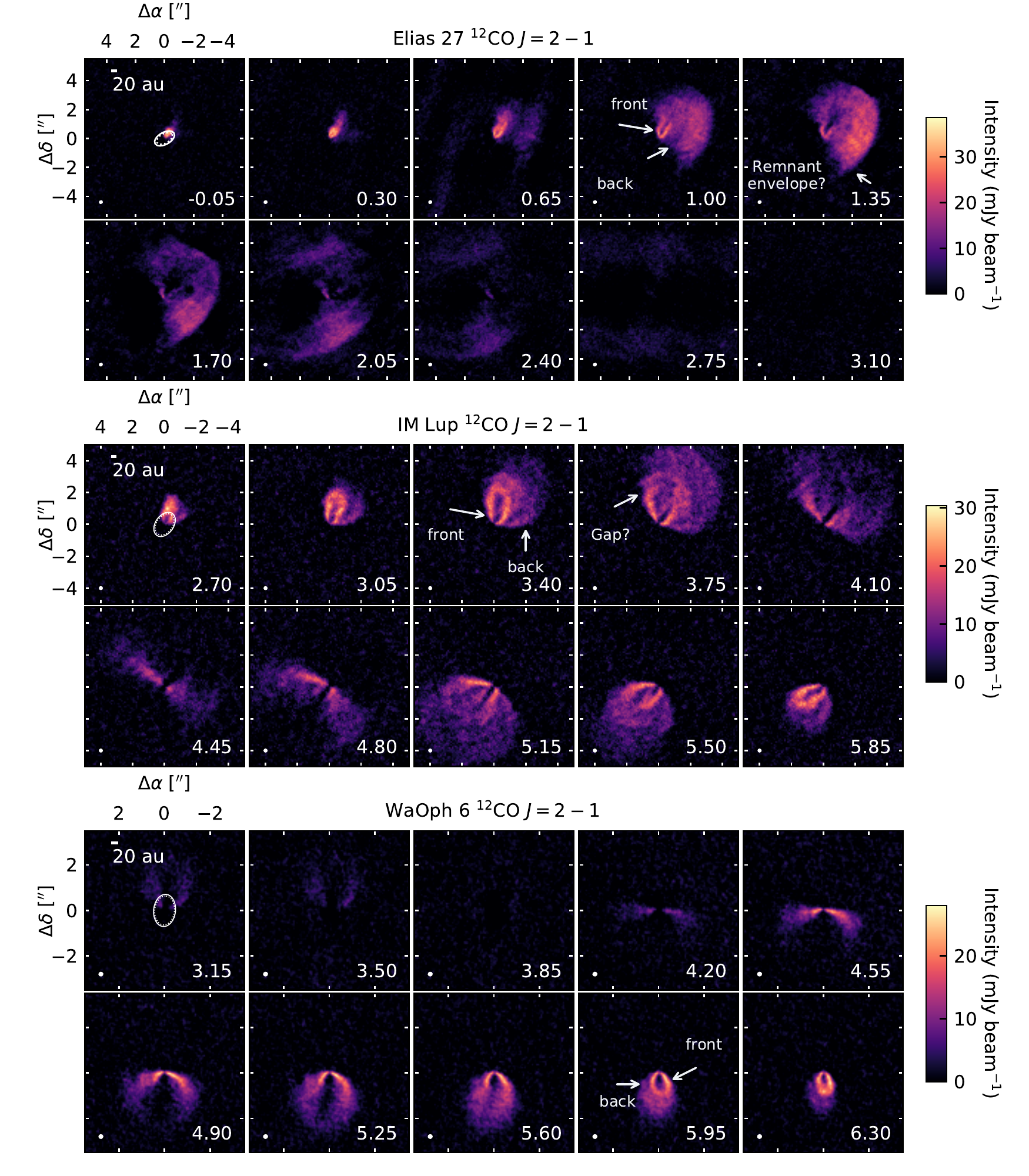}
\caption{Channel maps of the $^{12}$CO $J=2-1$ emission near the systemic velocity for the Elias 27, IM Lup, and WaOph 6 disks. The front and back sides of the disks, as well as some features of interest, are labeled with arrows. The positions of continuum annular substructures are marked with solid ellipses for rings and dotted ellipses for gaps in the first panel for each disk. The synthesized beam is shown in the lower left corner of each panel and the LSRK velocity in km s$^{-1}$ is printed in the lower right. See \citet{2018ApJ...869L..41A} for full channel maps. \label{fig:chanmaps}}

\end{figure*}

\begin{figure*}[htp]
\centering
\includegraphics[height = 2in]{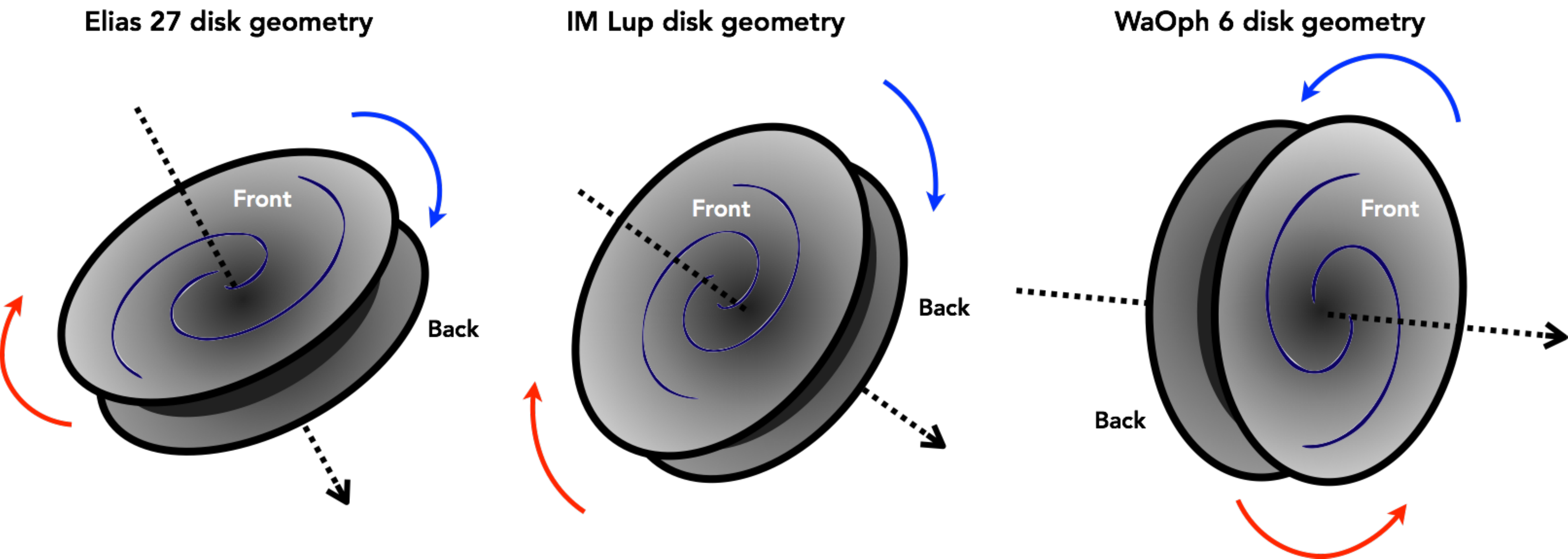}
\caption{Schematic showing the disk orientation and rotation direction for each source as constrained by $^{12}$ CO emission. The red and blue arrows show the directions from which redshifted and blueshifted $^{12}$CO emission emerge. The black arrow shows the rotation axis of the disk. The orientation of the continuum spiral arms is shown in purple. \label{fig:cartoons}}
\end{figure*}

The three sources have relatively high inclinations and visibly flared CO emission, allowing the brighter front side of the disk (i.e., the CO emitting layer in front of the midplane as viewed by the observer) to be differentiated from the dimmer back side (i.e., the CO emitting layer behind the midplane as viewed by the observer) \citep[e.g.,][]{2013ApJ...774...16R, 2018AA...609A..47P}. The terms ``front side'' and ``back side'' should not be confused with ``near side'' and ``far side,'' which are usually used to describe the relative orientation of the disk halves defined by the major axis of the projected image. Thus, we can establish whether the spiral arms are trailing (the outer ends point opposite to the direction of disk rotation) or leading (the outer ends point in the direction of disk rotation).

Channel maps of  $^{12}$CO $J=2-1$  are shown for each disk in Figure \ref{fig:chanmaps}. The emission geometry for Elias 27 indicates that the systemic velocity is $\sim2$ km s$^{-1}$, although this cannot be determined precisely because of the coarse velocity resolution and the severe foreground cloud absorption on the redshifted side. The extinction toward Elias 27 is high, with $A_V = 15$ \citep{luhman99b,2009ApJ...700.1502A}. The southwest side of the Elias 27 disk appears to be tilted toward the observer and the blueshifted emission originates northwest of the disk center, so the spiral arms must be trailing. The systemic velocity of IM Lup is $\sim4.5$ km s$^{-1}$ and the southwest side of the disk is also tilted toward the observer. The blueshifted emission originates northwest of the disk center and redshifted emission originates from the southeast, so the spiral arms are also trailing in this disk. Finally, the systemic velocity of the WaOph 6 disk is $\sim4.2$ km s$^{-1}$, and foreground cloud absorption is evident on the blueshifted side. The east side of WaOph 6 is tilted toward the observer. The blueshifted emission originates north of the disk center and the redshifted emission from the south, so the spiral arms also trail for WaOph 6. A diagram showing the disk orientations and rotation directions is presented in Figure \ref{fig:cartoons}. The orientations we find are consistent with that derived for Elias 27 from $^{12}$CO emission in \citet{2016Sci...353.1519P} and for IM Lup from $^{12}$CO emission in \citet{2018AA...609A..47P} and scattered light in \citet{2018ApJ...863...44A}. 

For all three sources, the non-axisymmetric residual plots (Figure \ref{fig:subtractedall}) exhibit a slight emission excess in the inner disk on the far side because the continuum peak is slightly offset from the disk center measured from fitting the annular substructures with ellipses \citep{2018ApJ...869L..42H}. These offsets are within the uncertainties established for the disk center positions (generally a few mas), but vertical structure may also contribute to the apparent brightness asymmetry. 

No obvious spiral structures are visible in $^{12}$CO emission for any of the three disks. However, the lines are optically thick and the angular resolution is about twice as coarse as that of the continuum images. Nevertheless, the line emission for IM Lup and Elias 27 have other possible features of interest. There are hints of ringlike substructures in IM Lup's $^{12}$CO emission at a distance of $\sim2\farcs5$ (400 au) from the disk center, which is outside the millimeter continuum emission. Because the observational setup was optimized for more compact continuum emission, additional observations with better $uv$ coverage should be obtained to confirm the molecular line substructures. Whereas IM Lup and WaOph 6 follow a Keplerian emission pattern, the emission in the Elias 27 disk shows evidence for non-Keplerian motion, particularly in the channel at 1.35 km s$^{-1}$. Optically thick line intensities typically peak toward the center of the disk due to higher temperatures, but the $^{12}$CO emission in the Elias 27 disk is broad and very bright (brightness temperatures exceeding 40 K) at a distance of a few arcseconds (several hundred au) from the disk center. While severe cloud contamination creates some ambiguity in interpreting the line data, the relative youth of this system suggests the possibility that this broad and bright component may be tracing remnant envelope material.

\begin{figure}[htp]
\centering
\includegraphics[scale = 1]{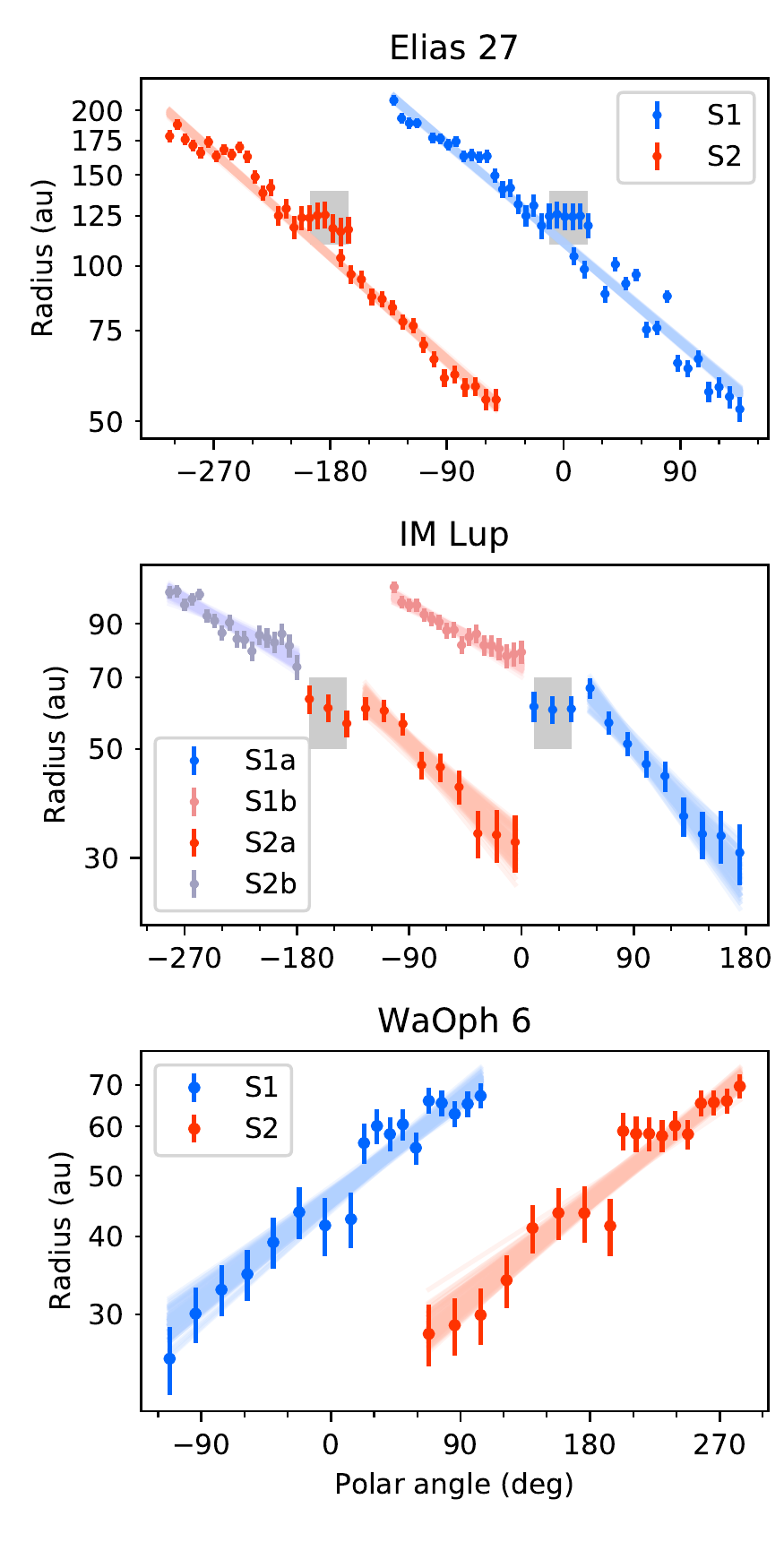}
\caption{Comparison of logarithmic spiral fits to the data. Measured spiral positions are plotted with $1\sigma$ error bars. Colored lines show logarithmic spirals derived from 100 random draws from the posterior for each arm. Gray boxes enclose points excluded from the fit. The y-axis is on a log-scale and S2 is plotted using unwrapped polar angle values for all sources. \label{fig:logspiralpolar}}
\end{figure}

\subsection{Spiral arm pitch angles\label{sec:pitchangles}}
To estimate the spiral arm pitch angles, we first measure the radial positions of the arms in each disk as a function of polar angle $\theta$ by deprojecting the nonaxisymmeric residual plots and searching for local maxima along rays of fixed $\theta$. This is similar to the approach used in \citet{2016Sci...353.1519P}. The uncertainties of the radial positions are assumed to be Gaussian and commensurate with the standard deviation of the beam. These uncertainties are scaled for deprojection, i.e., the uncertainties in $R$ at values of $\theta$ that fall near the minor axis of the projected disk are larger than at values of $\theta$ that fall near the major axis.

\begin{figure*}[htp]
\centering
\includegraphics[scale = 1]{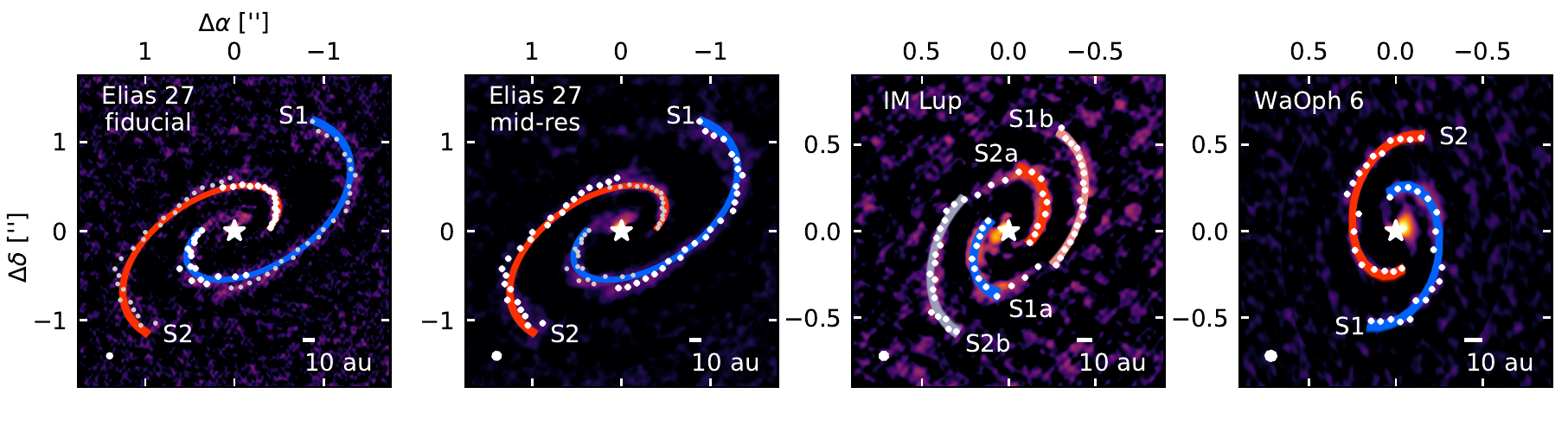}
\caption{Comparison of logarithmic spiral fits to the nonaxisymmetric residual maps. Measured spiral positions are marked as white dots. Colored curves show logarithmic spirals derived from 100 random posterior draws for each arm. For the Elias 27 disk, the points at $R<110$ au are measured from the high-resolution image and the points outside are measured from the mid-resolution image. For both the ``fiducial'' and ``mid-res'' images, the small gray points mark the positions measured from the other image. \label{fig:logspiralmaps}}
\end{figure*}

Because of the wide range of radii spanned by the spiral arms in the Elias 27 disk and the faint, broad emission at large radii, two separate images are used to measure the spiral arm positions in the inner and outer disk. The spiral arm ending on the northwest side of the disk is labeled ``S1'' and the arm ending on the southeast side is labeled ``S2.'' The arm positions are measured every 8$^\circ$ from $R = 50-110$ in a nonaxisymmetric residual map made from the fiducial image. The polar angle spacings for this and subsequent measurements are chosen such that the measured points are separated by about one synthesized beam. From $R = 110-230$ au, the arm positions are measured every 6$^\circ$ in the ``mid-res'' image, which was produced with a larger Gaussian taper in order to increase the SNR in the outer disk. The $R = 110$ au division is chosen based on where a slight radial depression appears in the nonaxisymmetric residuals in Figure \ref{fig:subtractedall}. To check for consistency, the positions identified in each image are compared to the emission morphology in the other image. We also re-imaged the other two disks with a larger taper to check that use of a slightly larger taper does not bias the measured positions of their spiral arms.

For the IM Lup disk, the spiral arm ending on the northwest side is labeled ``S1'' and the arm ending on the southeast side is labeled ``S2.'' The arm positions are measured every 15$^\circ$ between $R = 25-75$ au and then every 6$^\circ$ from $R=75-110$ au, with the radial breakpoint corresponding to the discontinuity in the nonaxisymmetric residual plot. The portions of the arms interior and exterior to $R = 75$ au are labeled ``a'' and ``b,'' respectively. 

For the WaOph 6 disk, the spiral arm ending on the south side is labeled ``S1'' and the arm ending on the north side is labeled ``S2.'' The arm positions are measured every 18$^\circ$ between $R = 25-50$ au and then every 9$^\circ$ from $R=50-80$ au, again with the radial breakpoint corresponding to the discontinuity in the nonaxisymmetric residual plot. 

 \begin{deluxetable*}{cccccc}
\tablecaption{Logarithmic spiral fit parameters\label{tab:spiralfits}}
\tablehead{
\colhead{Source} &\colhead{Spiral arm}&\colhead{Polar angle range measured\tablenotemark{a}}&\colhead{$R_0$}&\colhead{$b$}&\colhead{Pitch angle}\\
&&&(au)&&}
\startdata
Elias 27 & S1 &$-131^\circ$ to $136^\circ$&$110.9\pm0.6$&$-0.282\pm0.004$&$15\fdg7\pm0\fdg2$\\
& S2 &$56^\circ$ to $-52^\circ$ ($-304^\circ$ to $-52^\circ$)&$41.3\pm0.7$&$-0.295\pm0.004$&$16\fdg4\pm0\fdg2$\\
IM Lup & S1a &$55^\circ$ to $175^\circ$&$94\pm7$&$-0.40\pm0.04$&$22^\circ\pm2^\circ$\\
 & S1b &$-102^\circ$ to $0^\circ$&$74.6\pm1.6$&$-0.178\pm0.018$&$10^\circ\pm1^\circ$\\
 & S2a &$-125^\circ$ to $-5^\circ$&$30\pm2$&$-0.34\pm0.04$&$19^\circ\pm2^\circ$\\
& S2b &$78^\circ$ to $180^\circ$ ($-282^\circ$ to $-180^\circ$)&$43\pm3$&$-0.181\pm0.018$&$10^\circ\pm1^\circ$\\
WaOph 6 & S1 & $-112^\circ$ to $104^\circ$&$45.9\pm0.9$&$0.238\pm0.016$&$13\fdg4\pm0\fdg9$\\
& S2 &$68^\circ$ to $-76^\circ$ (68$^\circ$ to 284$^\circ$)&$21.3\substack{+1.4\\-1.3}$&$0.244\pm0.016$&$13\fdg7\pm0\fdg9$\\
\enddata
\tablenotetext{a}{If applicable, the phase-unwrapped polar angle range is given in parentheses.}
\end{deluxetable*}

The positions of the local maxima are plotted in Figure \ref{fig:logspiralpolar}. The plots show radius on a logarithmic scale versus polar angle on a linear scale because a spiral with constant pitch angle would appear as a straight line, and steeper slopes correspond to larger pitch angles. Some of these local maxima appear to trace constant or slowly varying radial positions near the radial depression at $R\sim110$ au from the Elias 27 nonaxisymmetric residuals and the discontinuity at $R\sim70$ from the IM Lup nonaxisymmetric residuals. If the intensity distribution at these radii is (nearly) axisymmetric (i.e., ringlike), then the local maxima positions from the residuals will be ``flattened'' against the ring because most of the emission at the ring itself will have been removed. These ``flattened'' regions are marked in Figure \ref{fig:logspiralpolar} and excluded from the spiral arm fits.  Near the radial discontinuity identified in the WaOph 6 disk at $\sim50$ au, a few of the measured positions also appear to ``flatten'' with radius, although it is less clear for this disk because fewer points are involved. 

We first measure the pitch angles under the simplest assumption that it is constant for a given arm. The positions $(\theta_i$, $R_i$) extracted for each arm are fit with a logarithmic spiral. The fitting procedure is described in Appendix \ref{sec:spiralfits}. The posterior medians and uncertainties computed from the 16th and 84th percentiles for $R_0$ and $b$, as well as the derived pitch angle, are listed in Table \ref{tab:spiralfits}. The logarithmic spiral fits are plotted in polar coordinates in Figure \ref{fig:logspiralpolar} and over the nonaxisymmetric residual maps in Figure \ref{fig:logspiralmaps}. For all disks, the arms appear to be symmetric. The estimated pitch angles in the Elias 27 disk are about twice as large as the pitch angles derived from logarithmic spiral fits in \citet{2016Sci...353.1519P}. The main source of this discrepancy is that \citet{2016Sci...353.1519P} subtracted a smoothed background profile from the disk emission in order to validate the presence of the annular gap (D69) in the residuals. However, a consequence of this approach was that the local maxima measured in the residuals were shifted outward from the spiral emission due to the general rise in emission that defines the outer boundary of the disk gap. By subtracting the median radial intensity profile instead from the original \citet{2016Sci...353.1519P} observations, we can recover pitch angles consistent with those measured from the new high resolution observations.

Logarithmic spirals may also not fully capture the spiral geometry; Figure \ref{fig:logspiralpolar} shows hints of a slightly decreasing pitch angle outside $R\sim150$ au. One difficulty in assessing whether the pitch angle is truly decreasing is that the spiral arms outside this radius are broad. Thus, even though the logarithmic spiral fits are slightly offset from the maxima identified in Figure \ref{fig:logspiralmaps}, they still lie well within the spiral arm emission. Logarithmic spirals also describe the WaOph 6 pattern reasonably well. Meanwhile, IM Lup exhibits a clear decrease in pitch angle between the inner and outer disk. The measured pitch angles for S1a and S2a are twice as large as those of S1b and S2b.

\begin{figure}[htp]
\centering
\includegraphics[scale = 1]{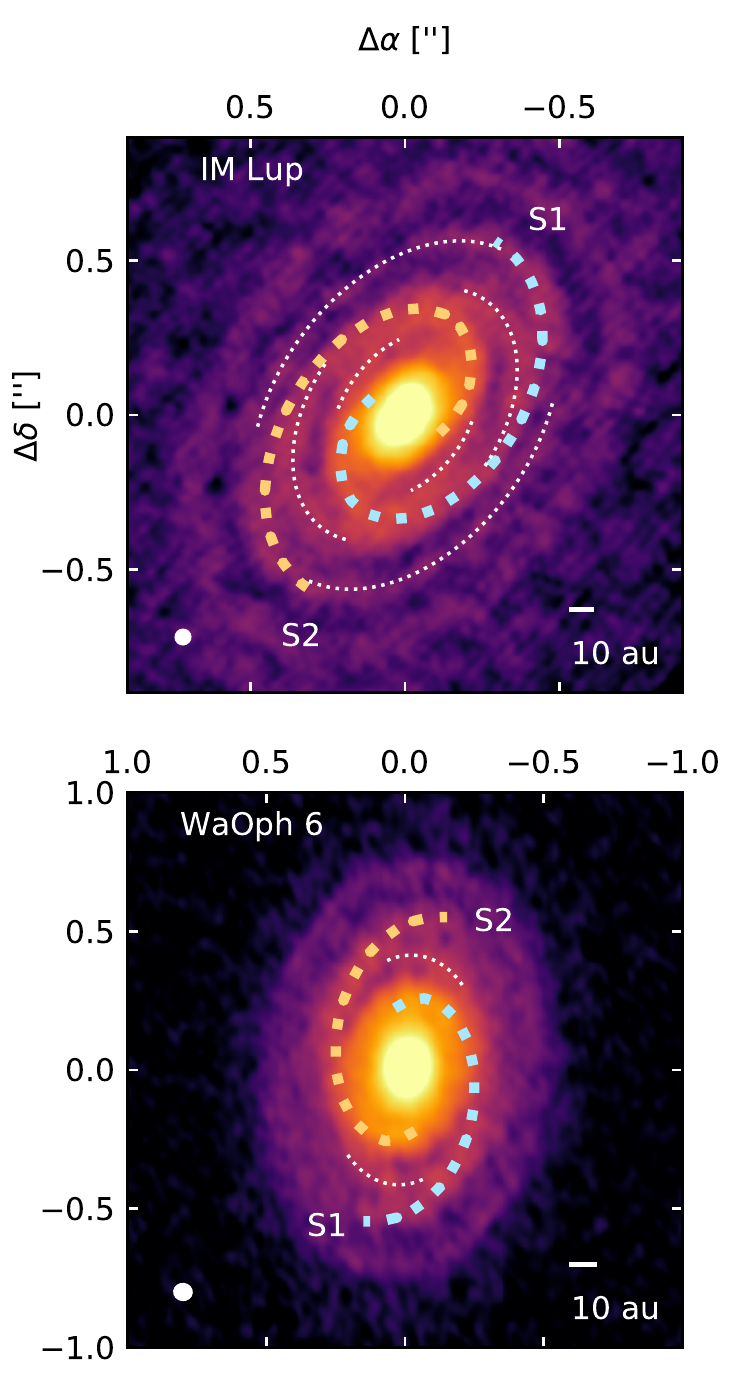}
\caption{\textit{Top}: Annotated continuum image of IM Lup. Best fit Archimedean spirals are plotted as blue and orange curves and radially constant structures at $R\sim98$ au are marked with white arcs to show the smooth connection to the spiral arms. Additional interarm structures are marked at $R\sim48$ and $R\sim72$ au. The color scale saturates at less than the peak intensity value to make fainter structures more visible \textit{Bottom}: Continuum image of WaOph 6 with best fit logarithmic spirals plotted as blue and orange curves and interarm structures marked with white arcs at $R\sim51$ au. \label{fig:possiblespurs}}
\end{figure}

The results of the logarithmic spiral fits motivate a comparison with a spiral arm parameterization in which pitch angle decreases with radius. To that end, we also fit the arm positions with Archimedean spirals, which take the form $R_m(\theta) = a+c\theta$. Although more complex expressions have been developed for the shapes of spiral wakes induced by companions in protoplanetary disks \cite[e.g.,][]{2002ApJ...569..997R, 2015ApJ...813...88Z}, we prefer the simpler Archimedean parametrization  because none of the spiral features are detected over more than a full winding, which would be key for discriminating more sensitively between different spiral forms. The fits are described in Appendix \ref{sec:archimedean}. 

In the Elias 27 disk, Archimedean spirals follow the shape of the arms in the outer disk better than the logarithmic spirals, but the pitch angles are too steep in the inner disk. Thus, while logarithmic and Archimedean spirals provide reasonable first-order approximations to the arm shapes in Elias 27, neither captures all of the structural nuances. The Archimedean spirals provide a simpler way to describe the pitch angle decrease in the IM Lup disk under the interpretation that the ``a'' and ``b'' components constitute a single pair of spiral arms. The spiral arms in WaOph 6 are also reasonably well-described by Archimedean spirals. 

However, both IM Lup and WaOph 6 appear to exhibit additional complexity beyond the two main arms identified. As shown in Figure \ref{fig:possiblespurs}, the structures traced by S1 and S2 in the IM Lup disk smoothly continue past the largest measured spiral positions. They are not noticeable in the nonaxisymmetric residual plot in Figure \ref{fig:subtractedall} because the structures become (nearly) radially constant, which suggests either that the pitch angles dramatically decrease past $R\sim98$ au or that the spiral arms merge with a ring structure. Additional interarm structures appear to branch off at $R\sim48$ and $R\sim72$ au. Because of the limited extent of the features, it is not clear whether these features belong to ring substructures intersecting the spiral structures or are additional spiral arms/branches. Like the IM Lup disk, the arms in WaOph 6 appear either to intersect with a ring structure or to have branches at $R\sim51$ au (see Figure \ref{fig:possiblespurs}). Better angular resolution will be needed to clarify the nature of these additional structures.

The finding that the spiral arms within each disk are symmetric is robust to the choice of spiral arm parametrization. Pitch angles are not well constrained in the inner 50 au of the disk unless the spiral arms are assumed to be logarithmic. Pitch angles derived from fitting logarithmic spirals are comparable to those derived in the outer disk ($R>50$ au) from fitting Archimedean spirals. The presence of additional ring substructure or spiral arm branching introduces some ambiguity in the pitch angle measurements, since radial substructure can create the appearance of a flattening pitch angle (and vice versa). 

\begin{figure*}[htp]
\centering
\includegraphics[width=\linewidth]{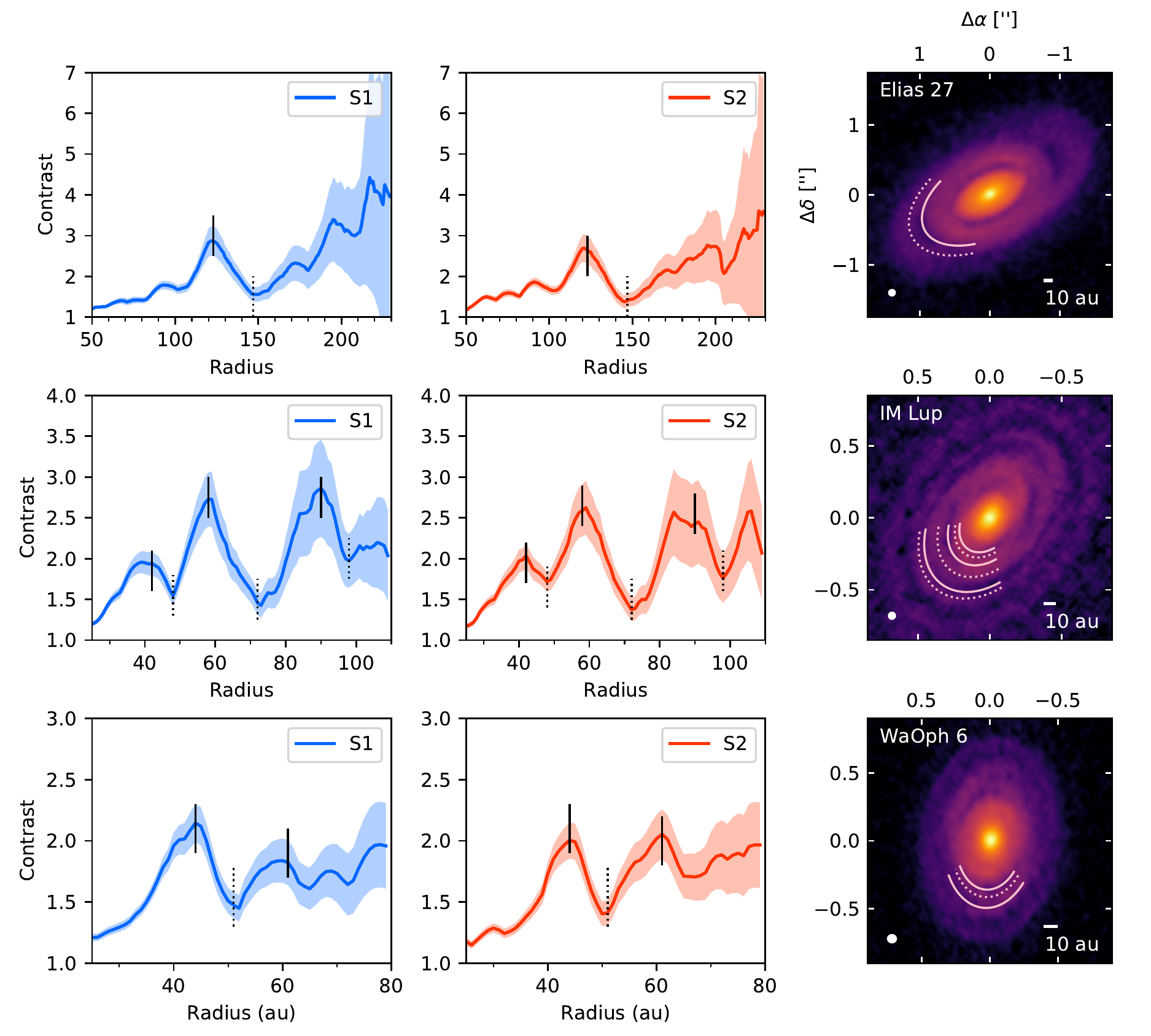}
\caption{Plots of spiral arm intensity contrasts.  \textit{Left column}: Intensity contrast of S1 as a function of radius for each disk. The shaded ribbon denotes the 1$\sigma$ uncertainty. Maxima in the contrast profiles are marked with solid vertical lines and minima are marked with dotted vertical lines. \textit{Middle}: Same as the left column, but for arm S2 in each disk. \textit{Right:} ALMA 1.25 mm continuum images marked at the radii corresponding to the contrast profile maxima (solid arcs) and minima (dotted arcs).\label{fig:contrasts}}
\end{figure*}

\subsection{Spiral arm contrasts}
To measure the spiral arm intensity contrasts for each disk, we first deproject the disk and measure the peak intensities of the spiral arms in radial bins that are one au wide. The ``background'' intensity is estimated by taking the fifth percentile of the pixel intensity distribution in each bin. The fifth percentile is chosen rather than the minimum because the former is more robust against outliers. Choosing slightly different percentile values to estimate the background leads to comparable contrast results. The contrasts are then calculated by taking the ratio of the peak intensities to the ``background,'' and the uncertainties of these ratios are calculated by assuming that both the maximum and background intensity uncertainties are defined by the image rms noise level and then propagating the error accordingly. For Elias 27 , the ``mid-res'' image ($\sim80$ mas beam) is used to estimate contrasts in order to have a sufficient SNR in the outer disk.

The spiral arm contrasts as a function of radius are plotted for each disk in Figure \ref{fig:contrasts}.  All three disks exhibit moderate contrasts, generally between 1.5 and 3. Low contrasts (i.e., close to 1) are observed at smaller radii, which may be a consequence of higher optical depths or genuinely lower surface density contrasts. 

In each disk, the individual arms have similar contrasts to one another, but show variations with radius. As previously shown in \citet{2016Sci...353.1519P}, the spiral arm contrast in the Elias 27 disk has a local maximum at $R\sim123$ au and a local minimum at $R\sim147$ au. (The radii stated in  \citet{2016Sci...353.1519P} are different because we use the new \textit{Gaia} parallax). The IM Lup disk exhibits many contrast variations with radius, but these must be interpreted carefully. Local minima are visible in the contrast profile at radii of $\sim$ 48, 72, and 98 au, which correspond to the locations where either the spiral arms are significantly decreasing in pitch angle or where additional interarm features are visible (see Figure \ref{fig:possiblespurs}). Similarly, the WaOph 6 contrast profile features a local minimum at $R\sim51$ au that also coincides with the presence of additional structure, as shown in Figure \ref{fig:possiblespurs}. One interpretation is that contrast levels are lower when spiral arms intersect with ringed structures. Alternatively, if arms are tightly wrapped near a given radius or spur structures are present, contrasts would be underestimated because beam smearing of the additional structures would increase the apparent  ``background'' emission level. 

The possibility of spatial filtering should be considered when measuring contrasts. For all three disks, the emission scales measured by the shortest baselines exceed the disk continuum sizes (see Table \ref{tab:observations}). Using the more conservative definition of maximum recoverable scale based on the 5th percentile of baseline lengths, the maximum recoverable scale for Elias 27 is smaller than the disk. However, the continuum flux measurements of these three disks from \citet{2018ApJ...869L..41A} are consistent with single-dish measurements from \citet{1994ApJ...420..837A} and \citet{2007AA...461..983V}, suggesting that the DSHARP observations are adequately recovering the continuum emission.
\section{Discussion \label{sec:discussion}}
\subsection{Comparison to other disks with spiral arms}

\subsubsection{Spiral arms in scattered light\label{sec:scattered}}
Spiral arms have been detected in scattered light in at least nine disks: AB Aur \citep{2004ApJ...605L..53F,2011ApJ...729L..17H}, DZ Cha \citep{2018AA...610A..13C}, HD 100453 \citep{2015ApJ...813L...2W, 2017AA...597A..42B}, HD 100546 \citep{2001AJ....122.3396G, 2013AA...560A..20B,2017AJ....153..264F}, HD 142527 \citep{2012ApJ...754L..31C, 2017AJ....154...33A}, LkH$\alpha$ 330 \citep{2016AJ....152..222A},  MWC 758 \citep{2013ApJ...762...48G,2015AA...578L...6B},  SAO 206462 \citep{2012ApJ...748L..22M, 2017ApJ...849..143S}, and V1247 Ori \citep{2016PASJ...68...53O}. Spiral arms have been tentatively identified for Oph IRS 48 \citep{2015ApJ...798..132F}, RY Lup \citep{2018AA...614A..88L}, and TW Hya \citep{2017ApJ...837..132V}. Finally, while \citet{2018ApJ...863...44A} classify the scattered light substructures in the IM Lup disk as concentric rings, they remark that some of those features may actually be tightly wound spirals. To our knowledge, scattered light images have not been published for either WaOph 6 or Elias 27.

Spiral arms in scattered light sometimes have $m=2$ symmetry similar to that seen in the DSHARP sources. However, so far there appears to be a morphological divide between the millimeter continuum of most disks known to have spiral arms in scattered light and the millimeter continuum morphologies of the three disks from this work. Most of the sources discussed in the previous paragraph have lopsided millimeter continuum morphologies and large emission cavities exceeding 20 au in radius \citep[e.g.,][]{2013ApJ...775...30I, 2013Sci...340.1199V, 2016ApJ...828...46A, 2017ApJ...840...32T,2017ApJ...848L..11K, 2018ApJ...860..124D,2018ApJ...864...81O, 2018AA...619A.161C, 2018arXiv181110365P}. On the other hand, Elias 27, WaOph 6, and IM Lup have symmetric spiral patterns, and the DSHARP spatial resolution ($\sim6-7$ au) is more than sufficient to rule out the presence of the cavities as large as those observed in the millimeter continuum of disks with scattered light spiral arms.

Because only a few of the disks known to have spiral arms in scattered light have been observed at millimeter wavelengths at an angular resolution comparable to that of DSHARP, it is not yet clear whether their high-contrast millimeter emission asymmetries are due to vortices, asymmetric spiral patterns, or some other origin. A high angular resolution millimeter continuum survey of the disks with spiral arms detected in scattered light would be useful for determining whether their morphologies still appear unified when resolved. If the asymmetries are unresolved spiral arms, the difference from the DSHARP disks may be a thermal effect, since spiral arm morphologies are strongly influenced by disk temperature \citep[e.g.,][]{2018ApJ...859..118B, 2018MNRAS.474L..32J}. The aforementioned asymmetric disks have stellar hosts with spectral types ranging from G to A, whereas IM Lup, WaOph 6, and Elias 27 are K or M stars. An age effect may also be at play if spiral arm morphologies change significantly over time. The disks with spiral arms detected in scattered light are thought to be older than the Elias 27, IM Lup, and WaOph disks by a few Myr \citep[e.g,][and references therein]{2018AA...620A..94G}, with the caveat that age estimates are highly uncertain. 

An interesting possibility is that the apparent spiral dichotomy is a consequence of fundamentally different spiral arm formation mechanisms. Hydrodynamical simulations have shown that sharp radial surface density gradients, such as those found at the edges of transition disk cavities, create favorable conditions for the Rossby wave instability to prompt the formation of vortices and spiral density waves \citep[e.g.,][]{1984MNRAS.208..721P, 1999ApJ...513..805L, 2001ApJ...551..874L}. This phenomenon has previously been invoked to explain the appearance of the SAO 206462 disk \citep{2016ApJ...819..134B, 2016ApJ...832..178V}. Given the absence of high-contrast asymmetries or prominent emission cavities in the Elias 27, IM Lup, and WaOph 6 disks, there is no obvious indication that the Rossby wave instability is operating in these sources. 

\subsubsection{Spiral arms in millimeter continuum emission}
Besides the three sources in this work, three other disks have been reported to have spiral features in millimeter continuum emission: MWC 758 \citep{2018ApJ...853..162B,2018ApJ...860..124D}, AS 205 N, and HT Lup A \citep{2018ApJ...869L..44K}. In contrast to disks with spiral arms observed in scattered light, disks with spiral arms observed at millimeter continuum wavelengths are predominantly hosted by K or M stars. This could be due to selection bias, since the majority of disks known to have millimeter continuum spiral arms were observed in DSHARP, which predominantly targeted K and M stars.

The five disks hosted by K and M stars (i.e., all except MWC 758) all have $m=2$ patterns. The case of MWC 758 is complicated\textemdash at least one spiral arm is detected on the southeast side of the disk and there are hints of another arm on the northwest side, although ambiguity is introduced by the disk's azimuthal asymmetries \citep{2018ApJ...853..162B,2018ApJ...860..124D}. Scattered light observations, though, reveal an $m=2$ spiral pattern in the upper layers of the disk \citep{2013ApJ...762...48G}. 

The dominance of $m=2$ spiral patterns so far could be due in part to the ambiguities involved in identifying single arm systems. For example, in millimeter continuum observations of V1247 Ori, \citet{2017ApJ...848L..11K} identify a ``bridge'' structure connecting an inner ring and outer crescent that appears to coincide with a spiral arm identified in scattered light, but stop short of classifying the structure itself as a spiral arm. 

One apparent difference between the spiral arms detected so far in multiple disk systems (AS 205 and HT Lup) and arms detected in systems without known companions (MWC 758, WaOph 6, IM Lup, and Elias 27) is that only the latter have annular substructures detected in conjunction with spiral arms. No obvious pattern emerges for the relative locations of the spiral and annular substructures, except that they occur in close proximity to one another.

\subsection{Possible origins of spiral structure}
\subsubsection{Spiral arms induced by a perturber}
Stellar and planetary companions are expected to trigger spiral density waves in protoplanetary disks \citep[e.g.,][]{1979ApJ...233..857G, 1986ApJ...307..395L, 2002ApJ...565.1257T}. A single companion can drive one or multiple arms \citep[e.g.,][]{2002MNRAS.330..950O,2018ApJ...859..118B}. Given the observed symmetry of the disks, though, it is unlikely that each arm is associated with a different companion. 

The spiral arm intensity contrasts measured for the disks in this work can be taken as an approximation of the surface density contrast if the azimuthal variations in temperature and dust opacity are small and the millimeter continuum emission is optically thin in the vicinity of the spirals. The assumption of azimuthally constant dust opacities is reasonable in the context of planet-disk interactions because the resulting spiral arms are not expected to trap dust particles \citep[e.g.,][]{2015MNRAS.451.1147J}. The optical depths outside $R>50$ au in the three disks also appear to be modest based on the analysis in \citet{2018ApJ...869L..42H}. Recent simulations have indicated that if a perturber were responsible for $m=2$ spiral patterns at the contrast levels measured in this work, it would have to orbit outside the arms and exceed several Jupiter masses \citep[e.g.,][]{2015ApJ...809L...5D,2015ApJ...813...88Z,2016ApJ...816L..12D,2017ApJ...839L..24M, 2018ApJ...859..119B}. If the spiral arms of the Elias 27, IM Lup, and WaOph 6 disks are indeed caused by stellar or planetary companions, then these massive objects in principle could be directly imaged in the near-infrared. 

To our knowledge, no stellar or planetary companions have been identified for Elias 27, IM Lup, or WaOph 6. Based on K-band contrasts from \citet{2005AA...437..611R} and K-band limits from \citet{2007MNRAS.379.1599L}, \citet{2017ApJ...839L..24M} derived a companion mass upper limit of 0.08 $M_\odot$ inside 350 au and 0.01 $M_\odot$ outside. In a SEEDS direct imaging survey of young stars that includes IM Lup, \citet{2017AJ....153..106U} estimated an upper limit of $\sim10$ $M_\text{Jup}$ within a few hundred au of the central star. A SPHERE survey of T Tauri disks that included IM Lup achieved a limit of $\approx$ 25 mag in H band and $\approx$ 25.5 mag in J band at a separation of $2''$ \citep{2018ApJ...863...44A}. Brown dwarfs and stars at 1 Myr are expected to be detectable at these limits \citep[e.g.][]{2015AA...577A..42B}, but \citet{2018ApJ...863...44A} do not identify any companion candidates within a few arcseconds of IM Lup.

Massive planets are also expected to create annular substructures in protoplanetary disks \citep[e.g.,][]{1984ApJ...285..818P, 2004AA...425L...9P, 2006Icar..181..587C,2010AA...518A..16F, 2015ApJ...809...93D}, and indeed, such structures are observed in all three disks. Based on their proximity, a relationship between the spiral arms and the annular structures in each disk might seem plausible.  However, simulations of planet-disk interactions find that the spiral arm pitch angle increases toward the location of a planet \citep[e.g.][]{2018ApJ...859..119B}, but pitch angles are not observed to increase toward gaps in the DSHARP disks. Moreover, \citet{2018ApJ...869L..47Z} point out that the high planet masses required to create large-scale symmetric spirals should also create much deeper gaps than what are actually observed. This might imply instead that any perturber inducing the spiral arms must be orbiting outside the detected extent of the millimeter continuum. Again, though, the pitch angles do not appear to be increasing with radius, which would be expected if the spirals were induced by wide-orbit objects. 
 
 An important caveat is that the observed millimeter continuum morphologies are being compared primarily to predictions for \textit{gas} morphologies. Most simulations of spiral arms in disks have focused on predicting morphologies in scattered light, which traces small grains that are well-coupled to the gas. The larger grains traced by millimeter continuum emission are not expected to be well-coupled to the gas, and thus millimeter continuum spirals may not have the same contrasts and geometries as gas spirals.  \citet{2018ApJ...860...27I} argue that because of this decoupling, planetary companions should not induce large-scale spiral arms at all in millimeter continuum emission. 

\subsubsection{Gravitational instability}
Gravitational instability (GI) is another oft-explored mechanism for forming spiral arms in circumstellar disks \citep[e.g.,][]{1998ApJ...503..923B, 2004MNRAS.351..630L,2008ApJ...681..375K, 2012ApJ...746..110Z}. Typically, Class II disks such as the DSHARP sources are thought to be too low-mass to drive global GI \citep[e.g.,][]{2016ARAA..54..271K}. In particular, previous estimates of the Toomre $Q$ parameter for the IM Lup and Elias 27 disks have indicated that they should be gravitationally stable \citep{2016Sci...353.1519P,2016ApJ...832..110C}. However, disk masses and temperatures are notoriously challenging to constrain for a number of reasons, including uncertainties related to dust opacity values, the dust size distribution, CO-to-H$_2$ conversation factors, dust-to-gas ratios, and high optical depths for continuum and line emission \citep[e.g.,][]{2016ApJ...828...46A, 2017AA...599A.113M,2017ApJ...844...99L,2017ApJ...841...39Y,2017MNRAS.470.1828E}. 

In theory, spiral pitch angles can be used to distinguish structures formed by companions from those formed via GI without needing to constrain the disk surface density and temperature profiles. GI can create a pair of symmetric, logarithmic spiral arms, while companions are expected to induce spiral arms with variable pitch angles \citep[e.g.,][]{2002ApJ...569..997R, 2015ApJ...813...88Z, 2015ApJ...812L..32D, 2018ApJ...860L...5F}. Several recent simulations of gravitationally unstable disks have reproduced the general morphology of the spiral arms in the Elias 27 disk \citep{2017ApJ...839L..24M, 2017ApJ...835L..11T, 2018MNRAS.477.1004H}. Excluding the discontinuity at $R\sim50$ au, the WaOph 6 spiral pattern can be approximated as a pair of logarithmic spirals. The IM Lup spirals appear symmetric but do not have constant pitch angles. On the other hand, the potentially branched structure of the IM Lup spirals is also seen in some simulations of gravitationally unstable disks \citep[e.g.,][]{2004ApJ...609.1045M, 2014MNRAS.444.1919D}. 

In practice, synthesized observations of spiral arms formed through the two mechanisms can appear similar \citep[e.g.,][]{2015ApJ...812L..32D, 2015ApJ...809L...5D, 2017ApJ...839L..24M}. In addition, as shown in this work, the presence of ringlike substructures or spur features significantly complicates analysis of pitch angles, and it can also be challenging to determine whether the pitch angle is increasing in the inner disk due to the angular resolution. The pitch angles derived in this work need to be treated with caution\textemdash any efforts to compare pitch angles derived from simulations to the observed pitch angles will need to account for the additional substructures. 

Another proposed way to distinguish between spiral arms induced by perturbers and those induced by GI is to search for evidence of particle trapping, since the former mechanism is not thought to lead to trapping \citep{2015MNRAS.451..974D}. This can be accomplished by observing the disks at multiple wavelengths and measuring the spectral index to determine whether there are spatial variations in the grain size distribution inside and outside the spiral arms. 

Aside from the spiral arms, several characteristics suggest that the disks in this work are among the Class II disks most likely to be gravitationally unstable. The Elias 27 and IM Lup disks are unusually large, with millimeter continuum emission stretching out to hundreds of au while most other disks have radial extents smaller than a few tens of au \citep[e.g.,][]{2017ApJ...845...44T, 2017AA...606A..88T, 2019MNRAS.482..698C}. This indicates that these two disks have unusually large surface densities in very cold regions, providing conditions hospitable for triggering GI. The WaOph 6 disk is much smaller than that of either Elias 27 or IM Lup, but it is still larger than the typical disk and has an SED that suggests it is particularly cold \citep{2009ApJ...700.1502A}. GI is also thought to be more likely to occur in younger sources \citep[e.g.,][]{2016ARAA..54..271K}.  Elias 27, IM Lup, and WaOph 6 are all estimated to be under 1 Myr old, placing them among the youngest sources in the DSHARP sample (again with the caveat that age estimates of pre-main sequence stars are highly uncertain).  Finally, hydrodynamical simulations indicate that gravitationally unstable disks are likely to have high stellar accretion rates, on the order of  $\dot M \sim10^{-6}$ $M_\odot$ yr$^{-1}$ \citep[e.g.,][]{2015ApJ...812L..32D,2015ApJ...805..115V}. The accretion rate of WaOph 6, at $\dot M =10^{-6.6\pm0.5}$ $M_\odot$ yr$^{-1}$ \citep{eisner05}, is close to this value. 
\subsubsection{Other hypotheses}
Shadowing from a misaligned inner disk has been proposed to trigger spiral arms observed in scattered light \citep{2016ApJ...823L...8M,2018MNRAS.475L..35M}, but the millimeter continuum observations of the disks in this work do not show evidence of misaligned inner disks. Furthermore, IM Lup has been observed in scattered light and does not exhibit signatures of shadowing \citep{2018ApJ...863...44A}. 

Stellar encounters may also create spiral arm structures in protoplanetary disks \citep[e.g.,][]{2003ApJ...592..986P, 2005AJ....129.2481Q}.  However, they are also expected to lead to substantial non-Keplerian motions in the disk, while the $^{12}$CO emission in the IM Lup and WaOph 6 disks appear to be largely Keplerian. The $^{12}$CO kinematics in the Elias 27 disk are suggestive of non-Keplerian motion, but do not exhibit any obvious tidal tails like those predicted by flyby simulations \citep[e.g.,][]{2015MNRAS.449.1996D}.

\subsection{How common are spiral arms?}
So far, it appears that only a minority of disks have millimeter continuum spiral arms. Whereas only six to date have confirmed millimeter continuum spiral structures, dozens of disks observed at moderate to high angular resolution are now known to have annular substructures \citep[e.g.,][and references therein]{2015ApJ...808L...3A, 2018ApJ...869L..41A, 2018ApJ...869L..42H, 2018arXiv181006044L}. Since most spiral-armed disks also have annular substructures, it is likely that the discrepancy between the occurrence rate of annular substructures and spiral arms will continue to hold as more disks are observed. 

The sources targeted for high angular resolution ALMA imaging have tended to be large and bright, and therefore probably have higher surface densities in cold outer disk regions compared to typical sources. If GI is the dominant mechanism for triggering spiral formation in disks around single stars, then the occurrence rate of spiral arms should decrease as fainter and smaller disks are observed. On the other hand, few large-cavity transition disks have been imaged at high angular resolution, even though they constitute 10 to 20\% of all protoplanetary disks \citep[e.g.,][]{2011ApJ...732...42A, 2016ApJ...828...46A}. Since many of the scattered light detections of spirals originate around transition disks (see the discussion in Section \ref{sec:scattered} and references therein), spiral structures might instead be underrepresented in existing observations. 

It is not clear whether spiral structures are less common than ringed structures because the spiral structures are shorter-lived, or because the conditions necessary to create observable spirals are intrinsically rarer. Reliably detecting spiral structures will require both high angular resolution and sensitivity, since the contrasts of spiral arms observed so far have been moderate, and insufficient resolution could easily lead to confusion between ringed/arclike structures and spiral arms with small pitch angles. Additional disk surveys that explore parts of parameter space different from DSHARP will clarify the circumstances under which spiral structures are likely to occur, which in turn will provide further insight into their origins. 

\section{Summary}\label{sec:summary}
The Disk Substructures at High Resolution Project has expanded the pool of disks with known spiral substructures in millimeter continuum emission from two to six. In this work, we analyze spiral structures in the Elias 27, IM Lup, and WaOph 6 disks. Our findings are as follows: 
\begin{enumerate}
\item The three disks all feature spiral patterns with $m=2$ symmetry. Determination of the absolute geometry of the disks from $^{12}$CO $J=2-1$ observations indicates that the spiral arms trail in all three disks. The spiral patterns are present throughout much of each disk and have azimuthal extents exceeding 200$^\circ$ in all cases. Spiral intensity contrasts are modest throughout, typically between 1.5 and 3. 

\item The structures of all three disks are remarkably complex. In all cases, annular substructures are present in addition to spiral structures. Most strikingly, the spiral arms in the Elias 27 disk  intersect a gap at $R\sim69$ au. Furthermore, while two main spiral arms are identified in all three disks, they have additional interarm structures that may be part of rings, branches from the main arms, or tightly wrapped continuations of the spiral arms. 

\item Unlike the millimeter continuum counterparts of many of the disks with spiral arms detected in scattered light, the millimeter continuum of the DSHARP disks does not exhibit high contrast, large-scale azimuthal asymmetries or large ($R>20$ au) emission cavities.  This apparent morphological divide may point to multiple spiral formation mechanisms operating in disks. Alternatively, this difference may be a thermal or age effect, since the DSHARP spiral-armed disks are generally young disks around K and M stars, while disks with spiral arms in scattered light are generally older and hosted by earlier-type stars. 

\item The pitch angles of the IM Lup spirals decrease with radius. The spiral arms of the Elias 27 and WaOph 6 disks can be approximated as logarithmic spirals (with constant pitch angle), although the measurement of pitch angle is complicated by additional substructures in the outer disk as well as large relative uncertainties in the spiral radial position in the inner disk. 

\item Previous numerical simulations indicate that large-scale symmetric spiral arms might be produced by stars or massive planets orbiting outside, but the observed spiral arms do not exhibit the predicted increase of pitch angle with radius. However, one limitation in comparing the observations to predicted spiral morphologies is that the latter primarily come from simulations of the gas distribution, which could be quite different from the large grain distribution traced by millimeter continuum emission. 

\item The observed spiral morphologies are reminiscent of numerical simulations of gravitationally unstable disks. The unusually large radial extents and relative youth of the disks studied in this work suggest that conditions may be favorable to GI. GI operating in these sources would imply that past analyses have either underestimated the disk surface densities or overestimated their temperatures.  
 
\end{enumerate}
\acknowledgments
This paper makes use of ALMA data\\
 \dataset[ADS/JAO.ALMA\#2016.1.00484.L]{https://almascience.nrao.edu/aq/?project\_code=2016.1.00484.L},\\
 \dataset[ADS/JAO.ALMA\#2013.1.00498.S]{https://almascience.nrao.edu/aq/?project\_code=2013.1.00498.S},\\
 \dataset[ADS/JAO.ALMA\#2013.1.00226.S]{https://almascience.nrao.edu/aq/?project\_code=2013.1.00226.S},\\
 \dataset[ADS/JAO.ALMA\#2013.1.00694.S]{https://almascience.nrao.edu/aq/?project\_code=2013.1.00694.S},\\
 and \dataset[ADS/JAO.ALMA\#2013.1.00798.S]{https://almascience.nrao.edu/aq/?project\_code=2013.1.00798.S}. We thank the NAASC and JAO staff for their advice on data calibration and reduction and Adam Kraus and Anjali Tripathi for helpful discussions. ALMA is a partnership of ESO (representing its member states), NSF (USA) and NINS (Japan), together with NRC (Canada) and NSC and ASIAA (Taiwan), in cooperation with the Republic of Chile. The Joint ALMA Observatory is operated by ESO, AUI/NRAO and NAOJ. The National Radio Astronomy Observatory is a facility of the National Science Foundation operated under cooperative agreement by Associated Universities, Inc. J.H. acknowledges support from the National Science Foundation Graduate Research Fellowship under Grant No. DGE-1144152. S.A. and J.H. acknowledge support from the National Aeronautics and Space Administration under grant No.~17-XRP17$\_$2-0012 issued through the Exoplanets Research Program.  C.P.D. acknowledges support by the German Science Foundation (DFG) Research Unit FOR 2634, grants DU 414/22-1 and DU 414/23-1. A.I. acknowledges support from the National Aeronautics and Space Administration under grant No.~NNX15AB06G issued through the Origins of Solar Systems program, and from the National Science Foundation under grant No.~AST-1715719.  L.P. acknowledges support from CONICYT project Basal AFB-170002 and from FCFM/U.~de Chile Fondo de Instalaci\'on Acad\'emica. V.V.G. and J.C acknowledge support from the National Aeronautics and Space Administration under grant No.~15XRP15$\_$20140 issued through the Exoplanets Research Program. Z.Z. and S.Z. acknowledge support from the National Aeronautics and Space Administration through the Astrophysics Theory Program with Grant No.~NNX17AK40G and the Sloan Research Fellowship. M.B. acknowledges funding from ANR of France under contract number ANR-16-CE31-0013 (Planet Forming disks). T.B. acknowledges funding from the European Research Council (ERC) under the European Union's Horizon 2020 research and innovation programme under grant agreement No.~714769.  L.R. acknowledges support from the ngVLA Community Studies program, coordinated by the National Radio Astronomy Observatory, which is a facility of the National Science Foundation operated under cooperative agreement by Associated Universities, Inc. 

\facilities{ALMA}

\software{ \texttt{analysisUtils} (\url{https://casaguides.nrao.edu/index.php/Analysis_Utilities}), \texttt{AstroPy} \citep{2013AA...558A..33A}, \texttt{CASA} \citep{2007ASPC..376..127M}, \texttt{emcee} \citep{2013PASP..125..306F}, \texttt{matplotlib} \citep{Hunter:2007}, \texttt{scikit-image} \citep{scikit-image}, \texttt{SciPy} \citep{scipy}}

\appendix

\section{Logarithmic spiral fits} \label{sec:spiralfits}
A logarithmic spiral takes the form $R(\theta) = R_0e^{b\theta}$, where $\theta$ is the polar angle in radians. The free parameters are $R_0$ and $b$. The pitch angle of a spiral is $\mu=\arctan\left| \frac{1}{r}\frac{dr}{d\theta}\right|$, so the pitch angle of a logarithmic spiral is $\mu = \arctan|b|$. Each arm is fit separately in order to assess whether the pitch angles vary between arms in the same disk. For the IM Lup disk, S1a, S2a, S1b, and S2b are fit independently because the S1b and S2b  positions have visibly shallower slopes on the radius-polar angle plot compared to S1a and S2a. For spiral arms that cross the $180^\circ/-180^\circ$ boundary, the polar angles are phase-unwrapped before fitting to eliminate the $360^\circ$ jump.

The log-likelihood takes the following form: 
\begin{equation}
\ln L = -\frac{1}{2}\sum_{i=1}^n \left[ \left(\frac{R_d(\theta_i)-R_m(\theta_i)}{\sigma_i}\right)^2+\ln(2\pi\sigma_i^2) \right],
\end{equation}
where $\theta_i$ is the polar angle, $R_d(\theta_i)$ is the radial position of the spiral arm measured at $\theta_i$, $R_m(\theta_i)$ is the model spiral arm radial position, and $\sigma_i$ is the uncertainty in the radial position. Flat priors are adopted for $R_0$ and $b$. The prior for $R_0$ is bounded by 0 and the disk dust radius $R_\text{dust}$ measured in \citet{2018ApJ...869L..42H}. The prior for $b$ is bounded by $-1$ and $0$ for Elias 27 and IM Lup and 0 and 1 for WaOph 6; the sign difference results from the different orientations of the spiral arms. The posterior probabilities are explored via the affine invariant MCMC sampler implemented in \texttt{emcee} \citep{2010CAMCS...5...65G, 2013PASP..125..306F}. The ensemble of 40 walkers is evolved for 15,000 steps. The first 500 steps are discarded as burn-in. Convergence of the MCMC chains is evaluated by measuring the autocorrelation time (generally less than 50 steps) and verifying that it is much smaller than the length of the chains.  

\begin{figure*}[htp]
\centering
\includegraphics[scale = 1]{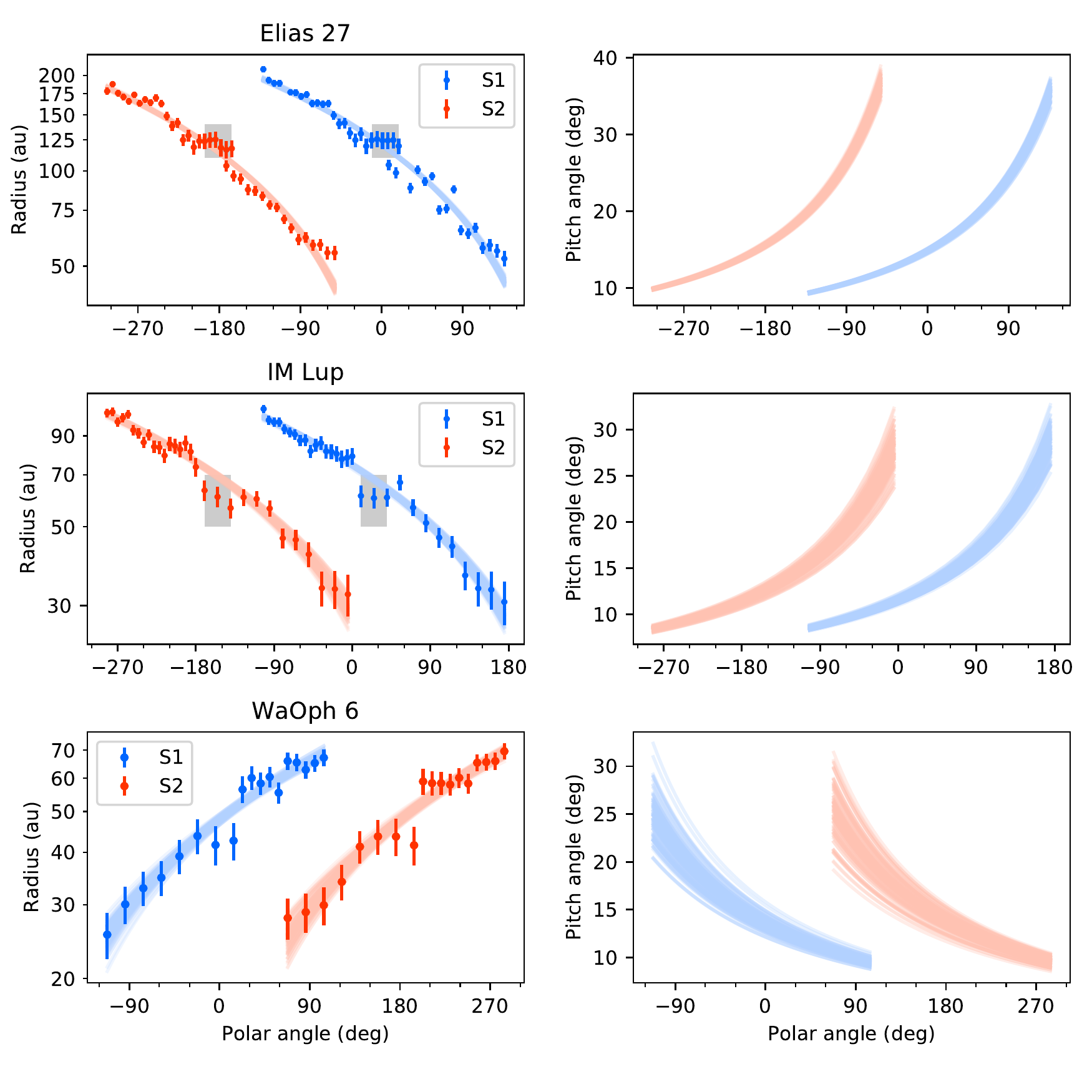}
\caption{\textit{Left}: Comparison of Archimedean spiral fits to the data. Measured spiral positions are plotted with $1\sigma$ error bars. Colored curves correspond to Archimedean spirals derived from 100 random draws from the posterior for each arm. Gray boxes enclose points excluded from the fit. The y-axis is on a log-scale and S2 is plotted using the unwrapped polar angle values for all sources. \textit{Right}: Pitch angles derived from 100 random draws from the posteriors, plotted as a function of polar angle.   \label{fig:archspiralpolar}}
\end{figure*}

\begin{figure*}[htp]
\centering
\includegraphics[scale = 1]{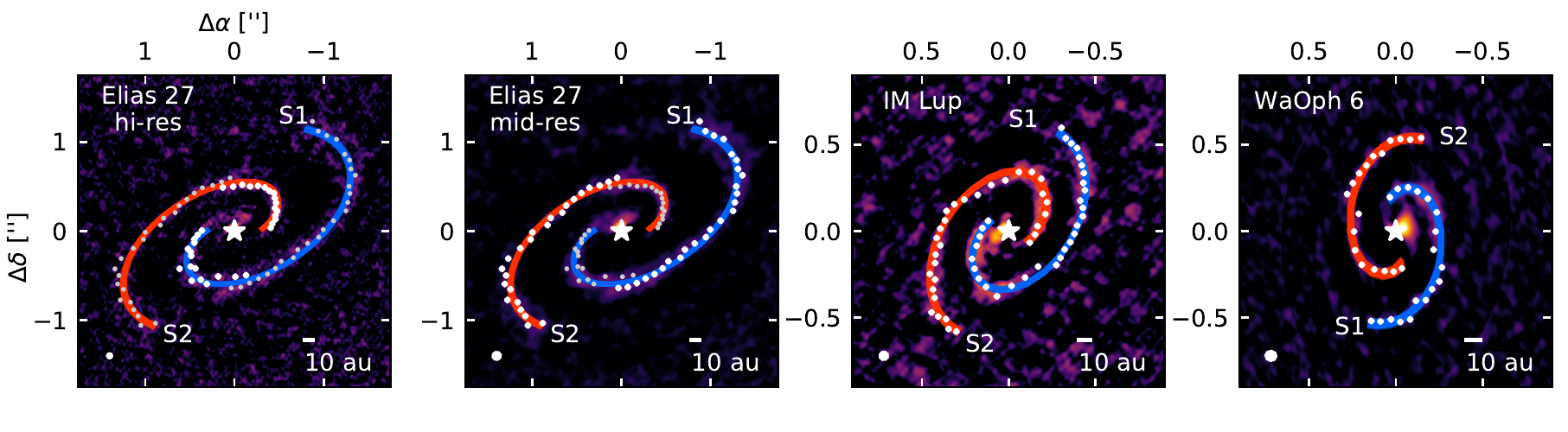}
\caption{Similar to Figure \ref{fig:logspiralmaps}, but for the Archimedean spiral fits.  \label{fig:archspiralmaps}}
\end{figure*}

\section{Archimedean spiral fits}\label{sec:archimedean}
The modeling is performed in a manner analogous to that of the logarithmic spiral fitting. The free parameters are $a$ and $c$. A flat prior is specified for $a$ over the range (0, $R_\text{dust}$). Likewise, a flat prior is specified for $c$ over the range (0, $R_\text{dust}$) for WaOph 6 and ($-R_\text{dust}$, 0) for the other two disks. The pitch angle is then calculated using the expression $\mu = \arctan\left|\frac{c}{R}\right|$. For the IM Lup disk, S1a and S1b are fit together as a single arm, as are S2a and S2b. The posterior medians and uncertainties computed from the 16th and 84th percentiles for $a$ and $c$ are listed in Table \ref{tab:archfits}. The Archimedean spiral fits and the derived pitch angles are shown as a function of $\theta$ in Figure \ref{fig:archspiralpolar}. Figure \ref{fig:archspiralmaps} show the Archimedean spiral fits over the nonaxisymmetric residual maps.

 \begin{deluxetable*}{ccccc}
\tablecaption{Archimedean spiral fit parameters\label{tab:archfits}}
\tablehead{
\colhead{Source} &\colhead{Spiral arm}&\colhead{Polar angle range measured\tablenotemark{a}}&\colhead{$a$}&\colhead{$c$}\\
&&&(au)&(au)}
\startdata
Elias 27 & S1 &$-131^\circ$ to $136^\circ$&$120.8\pm0.6$&$-32.1\pm0.4$\\
& S2 &$56^\circ$ to $-52^\circ$ ($-304^\circ$ to $-52^\circ$)&$14.1\pm1.3$&$-31.8\pm0.4$\\
IM Lup & S1 & $-102^\circ$ to $175^\circ$&$74.5\pm0.6$&$-15.2\pm0.4$\\
 & S2 & $78^\circ$ to $-5^\circ$ ($-282^\circ$ to $-5^\circ$)&$28.0\pm1.5$&$-15.3\pm0.4$\\
WaOph 6 & S1 & $-112^\circ$ to $104^\circ$&$47.8\pm0.8$&$11.5\pm0.7$\\
& S2 &$68^\circ$ to $-76^\circ$ (68$^\circ$ to 284$^\circ$)&$11\pm2$&$11.7\pm0.7$\\
\enddata
\tablenotetext{a}{If applicable, the phase-unwrapped polar angle range is given in parentheses.}
\end{deluxetable*}

\bibliographystyle{aasjournal}

\end{document}